\documentclass[10pt,a4paper,prc,notitlepage,nofootinbib,
showpacs,showkeys]{revtex4-1}
\usepackage[utf8]{inputenc}
\usepackage{amsmath}
\usepackage{amsfonts}
\usepackage{amssymb}
\usepackage{graphicx}
\usepackage{textcomp}
\usepackage{enumitem}
\usepackage{natbib}
\usepackage{hyperref}
\usepackage{comment}

\usepackage{tabularx}
\usepackage{tikz}
\usetikzlibrary{decorations.pathreplacing}
\usepackage{color}
\usepackage[figuresright]{rotating}
\usepackage[margin=1.25cm]{caption}
\usepackage[left=2cm,right=2cm,top=2cm,bottom=2cm]{geometry}
\begin{document}

\author{Jayson R.~Vavrek}
\author{Brian S.~Henderson}
\author{Areg Danagoulian\footnote{Corresponding author:  aregjan@mit.edu}}
\affiliation{Department of Nuclear Science and Engineering, Massachusetts Institute of Technology, Cambridge, MA, USA, 02139}



\title{Validation of Geant4's G4NRF module against nuclear resonance fluorescence data from $^{238}$U and $^{27}$Al} 



\begin{abstract}
G4NRF~\cite{jordan2007simulation,vavrek2018accuracy} is a simulation module for modeling nuclear resonance fluorescence (NRF) interactions in the Geant4 framework~\cite{allison2016_geant4}. In this work, we validate G4NRF against both absolute and relative measurements of three NRF interactions near 2.2~MeV in $^{238}$U and $^{27}$Al using the transmission NRF data from the experiments described in Ref.~\cite{vavrek2018experimental}. Agreement between the absolute NRF count rates observed in the data and predicted by extensive Geant4+G4NRF modeling validate the combined Geant4+G4NRF to within $15$--$20\%$ in the $^{238}$U NRF transitions and $8\%$ in $^{27}$Al, for an average $13\%$ discrepancy across the entire study. The difference between simulation and experiment in relative NRF rates, as expressed as ratios of count rates in various NRF lines, is found at the level of ${\lesssim}4\%$, and is statistically identical to zero. 
Inverting the analysis, approximate values of the absolute level widths and branching ratios for $^{238}$U and $^{27}$Al are also obtained.
\end{abstract}

\maketitle

\section{Introduction}
Nuclear resonance fluorescence (NRF)---the resonant absorption and re-emission of photons by a nucleus---is a nuclear measurement technique capable of providing a unique fingerprint for nearly any stable or long-lived isotope. Given its powerful isotopic discrimination, NRF has been used extensively as a probe of nuclear structure~\cite{kneissl1996investigation}. More recently, NRF has been proposed as a measurement technique for a variety of applied nuclear measurement scenarios such as nuclear warhead treaty verification~\cite{kemp2016physical, vavrek2018experimental}, spent fuel assay~\cite{quiter2011transmission}, and screening of cargo for contraband such as special nuclear material, explosives, drugs, and other controlled organic compounds~\cite{bertozzi2005nuclear, pruet2006detecting, kikuzawa2009nondestructive}.

In each of these measurements, the two key signatures from which all analysis ensues are the photon energies detected and the rate at which they are detected. The presence of characteristic spectral lines can be sufficient to confirm the presence of an isotope, while the photon detection rate in each spectral line is related to the isotope's concentration and NRF cross section. Measurements of relative NRF rates (e.g., rates normalized by a second NRF line or the $511$~keV peak) are often preferred in order to reduce systematic uncertainties---e.g., in the incident beam flux or detector efficiencies---that would make determinations of absolute NRF rates difficult. When possible, however, absolute NRF rate prediction is a potent tool for evaluating the feasibility of a proposed NRF experiment, and analysis of observed absolute NRF rates can provide a powerful absolute validation test of NRF modeling codes such as G4NRF~\cite{jordan2007simulation,vavrek2018accuracy} for Geant4~\cite{allison2016_geant4}. On its initial release, G4NRF was validated against absolute measurements only to within a factor of ${\sim}3$~\cite{warren_pers_comm}, and against relative measurements to within ${\sim}30\%$. An independent Geant4 NRF code by Lakshmanan et al.~\cite{lakshmanan2014simulations} has only been validated against relative measurements, and found a $20\%$ agreement in relative NRF rates. Our work improves the level of validation in our updated version of G4NRF to ${\sim}15\%$ agreement in absolute measurements and ${\lesssim}4\%$ agreement in relative measurements, the latter of which is statistically indistinguishable from zero.

Our previous study~\cite{vavrek2018accuracy} benchmarked G4NRF against semi-analytical modeling to within $1$--$3\%$. This study extends that effort by using experimental data to perform an experimental validation of G4NRF. In a recent work~\cite{vavrek2018experimental}, we conducted NRF measurements for a proof-of-concept nuclear warhead verification technique at the Massachusetts Institute of Technology (MIT) High Voltage Research Laboratory (HVRL). The experimental apparatus has been sufficiently well-characterized that we may further use the data from these $^{238}$U and $^{27}$Al NRF measurements for the dual purpose of performing improved absolute validation tests between the NRF data and a series of detailed Geant4+G4NRF models.

To this end, the structure of this paper is as follows: Section~\ref{sec:theory} describes the theory of nuclear resonance fluorescence interactions; Section~\ref{sec:experiments} introduces the $^{238}$U and $^{27}$Al NRF experiments \cite{vavrek2018experimental} that produced the data used in this work; Section~\ref{sec:predictions} describes the Monte Carlo NRF simulation models using the G4NRF package for the Geant4 toolkit; Section~\ref{sec:results} compares the data to the simulations; finally Section~\ref{sec:discussion} concludes with a discussion of results, systematics, and prospects for future work on NRF model and data validation.

\section{Theory}\label{sec:theory}
Nuclear resonance fluorescence is the photonuclear interaction in which a nucleus with a resonant energy level $E_r$ absorbs a photon of energy $E \simeq E_r$, promoting the nucleus to the excited state at $E_r$. The excitation modes are typically M1, E1, or E2~\cite{kneissl1996investigation}. In particular, $^{238}$U transitions near $E_r=2$--$3$~MeV correspond to the M1 `scissors modes', in which groups of neutrons and protons (`orbital scissors') or groups of spin-up and spin-down nucleons (`spin scissors') oscillate against each other~\cite{balbutsev2018scissors, kneissl1996investigation, heyde2010magnetic}. The excited state then decays with a lifetime $\mathcal{O}(\text{fs})$ to a lower energy level $E_j$ (often the ground state, $E_0 \equiv 0$), emitting a photon of energy $E' \simeq E - E_j$ with branching ratio $b_{r,j}$. Summed over all decay modes $j$, the cross section for NRF absorption through a resonant energy level $E_r$ is a Doppler-broadened Breit-Wigner distribution:\footnote{For brevity, we include only the most salient equations and definitions, and refer the reader to Refs.~\cite{vavrek2018accuracy, metzger1959resonance, kneissl1996investigation} for full detail.}
\begin{align}\label{eq:sigmaNRF}
\sigma_r^\text{NRF}(E) = 2\pi^{1/2} g_r \left( \frac{\hbar c}{E_r} \right)^2 \frac{b_{r,0}}{\sqrt{t}} \int_{-\infty}^{+\infty} \exp\left[ -\frac{(x-y)^2}{4t} \right] \frac{dy}{1+y^2},
\end{align}
where $g_r \equiv (2J_r+1)/(2(2J_0+1))$ is a spin statistical factor determined by the resonant level and ground state spins $J_r$ and $J_0$, $b_{r,0}$ is the branching ratio between $E_r$ and the ground state, and $t$ and $x$ involve the natural and Doppler-broadened widths ($\Gamma_r$ and $\Delta$, respectively) of the level corresponding to $E_r$:
\begin{align}
x &\equiv 2(E-E_r)/\Gamma_r\\
t &\equiv (\Delta/\Gamma_r)^2\\
\Delta &\equiv E \sqrt{\frac{2k_BT}{Mc^2}}
\end{align}
with $T$ the absolute temperature and $Mc^2$ the rest-mass energy of the target nucleus.

In the literature, the `integrated cross section' is often used as an approximate measure of the strength of a resonance:
\begin{align}\label{eq:sigma_int}
\sigma_r^\text{int} = \int \sigma_r^\text{NRF}(E)\, dE = 2\pi^2 g_r \left( \frac{\hbar c}{E_r} \right)^2 \Gamma_{r,0},
\end{align}
assuming $E_r \gg \Delta \gg \Gamma_r$. In the thin-target limit, the expected NRF count rate is directly proportional to $\sigma_r^\text{int}$, eliminating any dependence on the true shape of the cross section (Eq.~\ref{eq:sigmaNRF})~\cite{vavrek2018accuracy}. The thin-target limit does not hold for the targets (foil and measurement objects) used in this work, and the accuracy goal is sufficiently strict, such that numerical integration of Eq.~\ref{eq:sigmaNRF} is required for high-accuracy NRF count rate predictions.\footnote{Even the Gaussian approximation common in the literature (see, e.g., Ref.~\cite{vavrek2018accuracy}) will induce errors as large as $6\%$ in these geometries.}

The angular dependence of NRF photon emission is given by the angular differential cross section,
\begin{align}
\frac{d\sigma_r^\text{NRF}(E)}{d\Omega} = \frac{W(\theta, \phi)}{4\pi} \sigma_r^\text{NRF}(E),
\end{align}
where $W(\theta, \phi)$ is known as the angular correlation function for emission into the solid angle $d\Omega = d \cos\theta d\phi$. For unpolarized beams, the $\phi$-dependence disappears, and for the $0\to 1\to 0$ spin sequence in the $^{238}$U ground state transitions studied in this work, the $W(\theta, \phi)$ are
\begin{align}
W_\text{010} (\theta, \phi) = \frac{3}{4} \left( 1 +  \cos^2\theta \right).
\end{align}
Ref.~\cite{hamilton1940directional} tabulates the $W(\theta, \phi)$ for various other spin sequences.

\section{Absolute NRF rate experiments}\label{sec:experiments}
Although the experiments of Ref.~\cite{vavrek2018experimental} were designed for \textit{relative} measurements of depleted uranium (DU) and Al NRF signatures, they can still be used for \textit{absolute} measurements of the NRF count rates given good understanding of the experimental design. To focus the analysis, we will follow Refs.~\cite{vavrek2018experimental, vavrek2018accuracy} and examine only the $^{238}$U NRF lines at 2.176 and 2.245~MeV and the $^{27}$Al line at 2.212~MeV.

\subsection{Experimental design}
\begin{figure}
\centering
\begin{tikzpicture}
\fill[fill=gray] (0,-0.5) rectangle (2,0.5);
\node[below] at (1,-0.5) {collimator};
\fill[fill=white, draw=white] (0,-0.05) -- (0,0.05) -- (2,0.15) -- (2,-0.15);


\fill[fill=brown!80!black] (-0.5, -0.5) rectangle (-0.25, 0.5);
\fill[fill=yellow!80!black] (-0.5, -0.2) rectangle (-0.4, 0.2);
\node[above] at (-0.375,0.5) {radiator};
\draw (-0.375, -0.6) -- (-0.45, -1.4);
\draw (-0.45, -0.22) -- (-1.0, -1.4);
\node[below] at (-0.45,-1.4) {Cu};
\node[below] at (-1.0,-1.4) {Au};
\draw (-0.26, -0.40) -- (-0.26, -1.0) -- (0.25, -1.0) -- (0.25, -2.0);
\draw (0.25,-2.3) circle (0.3);
\node at (0.25, -2.3) {A};
\draw (0.25, -2.6) -- (0.25, -3.2);
\draw (0.10, -3.2) -- (0.40, -3.2);
\draw (0.15, -3.25) -- (0.35, -3.25);
\draw (0.20, -3.3) -- (0.30, -3.3);

\fill[fill=brown!80!black] (3,-1) rectangle (3.2,1);
\fill[fill=gray!60!white] (3.2,-0.8) rectangle (3.25,0.8); 
\fill[fill=blue!80!black] (3.25,-0.8) rectangle (3.45,0.8);
\draw [decorate,decoration={brace,amplitude=3pt}](3.0,1.2) -- (3.7,1.2) 
node[yshift=12pt,xshift=-8pt]{measurement object}; 
this
\fill[fill=gray!60!white] (3.45,-0.8) rectangle (3.5,0.8);
\fill[fill=brown!80!black] (3.5,-1) rectangle (3.7,1);
\node[below] at (2.7,-1.4) {plastic};
\node[below] at (4.1,-1.4) {DU, Al};
\draw (3.1,-1.1) -- (2.9,-1.4);
\draw (3.6,-1.05) -- (2.9,-1.4);
\draw (3.35,-1.) -- (3.7,-1.4);

\fill[fill=blue!80!black] (9.8,-0.8) rectangle (10,0.8);
\fill[fill=gray!60!white] (10,-1) rectangle (10.4,1);
\draw [decorate,decoration={brace,amplitude=3pt}]
(9.8,1.1) -- (10.4,1.1) node[xshift=10pt,yshift=10pt]
{DU/Al foil};

\fill[fill=gray] (13.75,-0.30) rectangle (15.0,0.30);
\fill[fill=black] (14,-0.15) rectangle (14.75,0.15);
\node[above] at (14.375,0.30) {LaBr$_3$};

\fill[fill=gray, rotate around={55:(10.3,0)}] (5.5,-0.65) rectangle (7.4,2.65);
\fill[fill=gray, rotate around={-55:(10.3,0)}] (5.5,-2.65) rectangle (7.4,0.65);
\node at (6.7,2.3) {Pb};
\node at (6.7,-2.3) {Pb};

\fill[fill=black, rotate around={55:(10.3,0)}] (5.7,-0.35) rectangle (7.2,0.35);
\fill[fill=black, rotate around={-55:(10.3,0)}] (5.7,-0.35) rectangle 
(7.2,0.35);
\node at (10,3.5) {2 $\times$ HPGe};
\node at (10,-3.5) {1 $\times$ HPGe};
\draw (8.4,3.5) -- (9.1,3.5);
\draw (8.4,-3.5) -- (9.1,-3.5);

\draw[dashed] (-1,0) -- (15.2, 0);

\draw[-{latex}, green!70!black, line width=1pt] (-0.2,0) -- (2.95,  0.15);
\draw[-{latex}, green!70!black, line width=1pt] (-0.2,0) -- (2.95, -0.15);
\draw[-{latex}, green!70!black, line width=1pt] (-0.2,0) -- (2.95, 0);
\draw (2.6, 0.35) -- (2.3, 0.6) -- (1.4, 0.80);
\node[above] at (1.4, 0.75) {$\phi_0(E)$};

\draw[-{latex}, green!70!black, line width=1pt] (3.75, 0.1) -- (9.75, 0.2);
\draw[-{latex}, green!70!black, line width=1pt] (3.75, -0.1) -- (9.75, -0.2);
\node[above] at (5, 0.2) {$\phi_t(E)$};

\draw[dashed, rotate around={55:(10.3,0)}] (10.3,0) -- (7.7, 0);
\draw[dashed] (10.8,0) arc (360:235:0.5cm);
\node at (11.5,-0.5) {$\theta_d \simeq 125^\circ$};
\draw[-{latex}, green!70!black, line width=1pt, rotate around={-55:(10.3,0)}] 
(9, 0) -- (7.5, 0);
\draw[-{latex}, green!70!black, line width=1pt, rotate around={55:(10.3,0)}] (9, 
0) -- (7.5, 0);
\node at (9.6,2.2) {NRF $\gamma$};
\node at (9.6,-2.2) {NRF $\gamma$};

\draw[-{latex}, red!70!black, line width=1pt] (-1,0) -- (-0.45, 0);
\node[below] at (-1, 0) {e$^-$};

\draw[-{latex}, black, line width=0.5pt] (0,3) -- (0,2);
\node[below] at (0,2) {$+\hat{z}$};
\draw[-{latex}, black, line width=0.5pt] (0,3) -- (1,3);
\node[right] at (1,3) {$+\hat{x}$};
\draw (0,3) circle (0.1);
\fill[black] (0,3) circle (0.05);
\node[above] at (0,3) {$+\hat{y}$};

\node at (13.0, -3.8) {(top view, not to scale)};

\end{tikzpicture}
\caption{Schematic top view of the HVRL experiment, adapted from Ref.~\cite{vavrek2018experimental}. 
}
\label{fig:hvrl_schematic}
\end{figure}
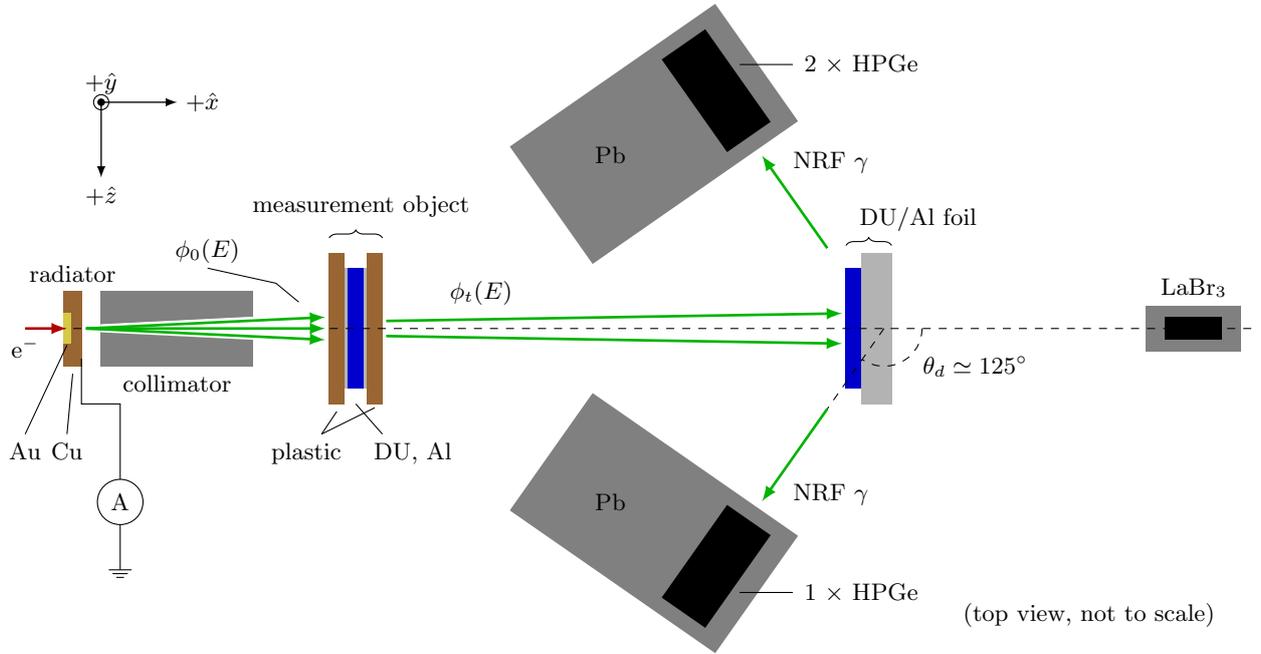

\begin{figure}
\centering
\includegraphics[width=0.75\columnwidth]{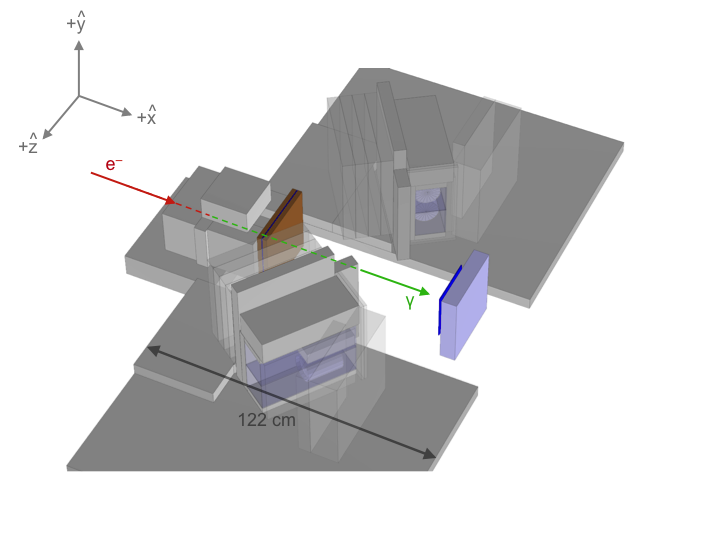}
\caption{Detailed Geant4 model of the HVRL geometry (to scale). The measurement object (here plastic and DU) is shown on the left in brown and blue, while the DU and Al foil components are shown on the right in dark and light blue, respectively. 
}
\label{fig:g4geom}
\end{figure}

In an experimental setup at the MIT High Voltage Research Laboratory (HVRL)---see Figs.~\ref{fig:hvrl_schematic} and \ref{fig:g4geom}---a ${\sim}25$~{\textmu}A beam of electrons was accelerated to a kinetic energy of $2.52$~MeV using a van de Graaff generator.\footnote{Diagnostics used to determine and monitor the bremsstrahlung beam parameters, including beam centering and spot size, are given in the SI Appendix of Ref.~\cite{vavrek2018experimental}. A report on preliminary experiments can be found in Ref.~\cite{vavrek2017progress}.} The beam impinged on a radiator constructed of $126$~{\textmu}m of Au followed by ${\sim}1$~cm of Cu, producing a bremsstrahlung photon spectrum with an energy endpoint of $2.52$~MeV. The bremsstrahlung interrogation beam then passed through a 20~cm-long conical collimator of entry diameter 9.86~mm and exit diameter of 26.72~mm, producing a maximum spread of photon angles of approximately $5^\circ$.

The collimated beam ($\phi_0(E)$ in Fig.~\ref{fig:hvrl_schematic} and Section~\ref{sec:theory}) then impinged on one of various measurement objects, which were designed as targets for a proof-of-concept demonstration of a warhead verification protocol using NRF~\cite{vavrek2018experimental}. These objects (namely, Objects~1 and 3 in Table~\ref{tab:geometries}) therefore consisted of DU plates as a proxy for a spheroidal fissile core and high-density plastic plates as a proxy for the shell of conventional explosives. In Objects~2 and 4 of Table~\ref{tab:geometries}, the DU was replaced with a similar areal density of Pb to test the NRF measurement's sensitivity to isotopic changes that would be difficult to detect through isotope-insensitive measurement techniques such as simple radiography. In Object~5, only half the DU was replaced with Pb.

\begin{table}[!h]
\begin{tabular}{ccc}
name & metal plate thicknesses & total areal density [g/cm$^2$]\\\hline
foil & 3.28~mm DU, 63.5~mm Al & 23.4\\
Object~1 & 3.72~mm DU & 12.1\\
Object~2 & 5.29~mm Pb & 11.0\\
Object~3 & 7.19~mm DU & 18.7\\
Object~4 & 10.58~mm Pb & 17.0\\
Object~5 & 3.72~mm DU, 5.29~mm Pb & 18.1\\
\end{tabular}
\caption{Compositions of the foil and measurement objects. In Objects~1--5, the DU and Pb plates are located between two additional 19~mm high-density polyethylene layers.}
\label{tab:geometries}
\end{table}

The flux transmitted through the object ($\phi_t(E)$ in Fig.~\ref{fig:hvrl_schematic} and Section~\ref{sec:theory}) then struck the DU+Al foil, which was constructed from 3.28~mm of DU followed by 63.5~mm of standard density Al. The radius of the beam spot on the DU plate was approximately 52~mm, with only a small amount of illumination outside this radius due to scatter in the collimator. The distance from the collimator output to the DU plates was about 76~cm, or about 1~m from the Au radiator to the DU plates; the separation between the DU and Al plates (due to the base of the DU stand) was 2.5~cm. Three $100\%$ relative efficiency ORTEC GEM high-purity germanium (HPGe) detectors were placed ${\sim}55$~cm from the center of the DU component of the foil, at angles of $\theta_d \simeq 125^\circ$ to the beamline, in order to record the NRF spectra emitted by the foil. Data acquisition (DAQ) from the HPGe detectors was accomplished using Canberra Lynx Digital Signal Analyzers, which were controlled by the custom-written Python Read-Out with Lynx for Physical Cryptography (PROLyPhyC) software based on the Lynx Software Development Kit. Spectral data were acquired in 32768-channel histogram mode, in intervals of five minutes (real time) so as to reduce the likelihood of data corruption by beam instabilities.

To reduce the flux of active background photons (primarily at low energies) that would induce pileup and detector deadtime, the detectors were shielded with significant amounts of Pb, ranging from around 5~cm below the detectors to 25--30~cm in the direction of the measurement object and radiator. Between the foil and the HPGe detectors, only a 2.54~cm-thick lead filter was present (located 5--8~cm from the front of each detector casing) in order to reduce the low-energy flux without unduly attenuating the NRF flux. A depiction of the Pb shielding is given in Fig.~\ref{fig:g4geom}.

A 38.1~mm right square cylinder LaBr$_3$ scintillator detector was also placed several meters behind the target foil as an independent measurement of the bremsstrahlung beam flux transmitted through the measurement object and foil. The photon energy deposition in the scintillator crystal was recorded, again at five-minute intervals, using a CAEN DT-5790M digitizer controlled through the ADAQAcquisition~\cite{hartwig2016adaq} software.

The electron current was measured on the bremsstrahlung radiator itself with a Keithley Model~614 Electrometer. The analogue output of the electrometer was digitized at a rate of 1~kHz using a Measurement Computing Model~USB-201 ADC, and the average current over the course of each acquisition period was subsequently computed. The product of the average beam current with the detector live time then gives the `live charge' delivered during the run, so that the observed spectra---and thus the rate of NRF photon detections---can be normalized by the number of electrons incident on the radiator while the detectors were live.

\subsection{Analysis of experimental data}
\begin{figure}[!htb]
\centering
\includegraphics[width=0.8\columnwidth]{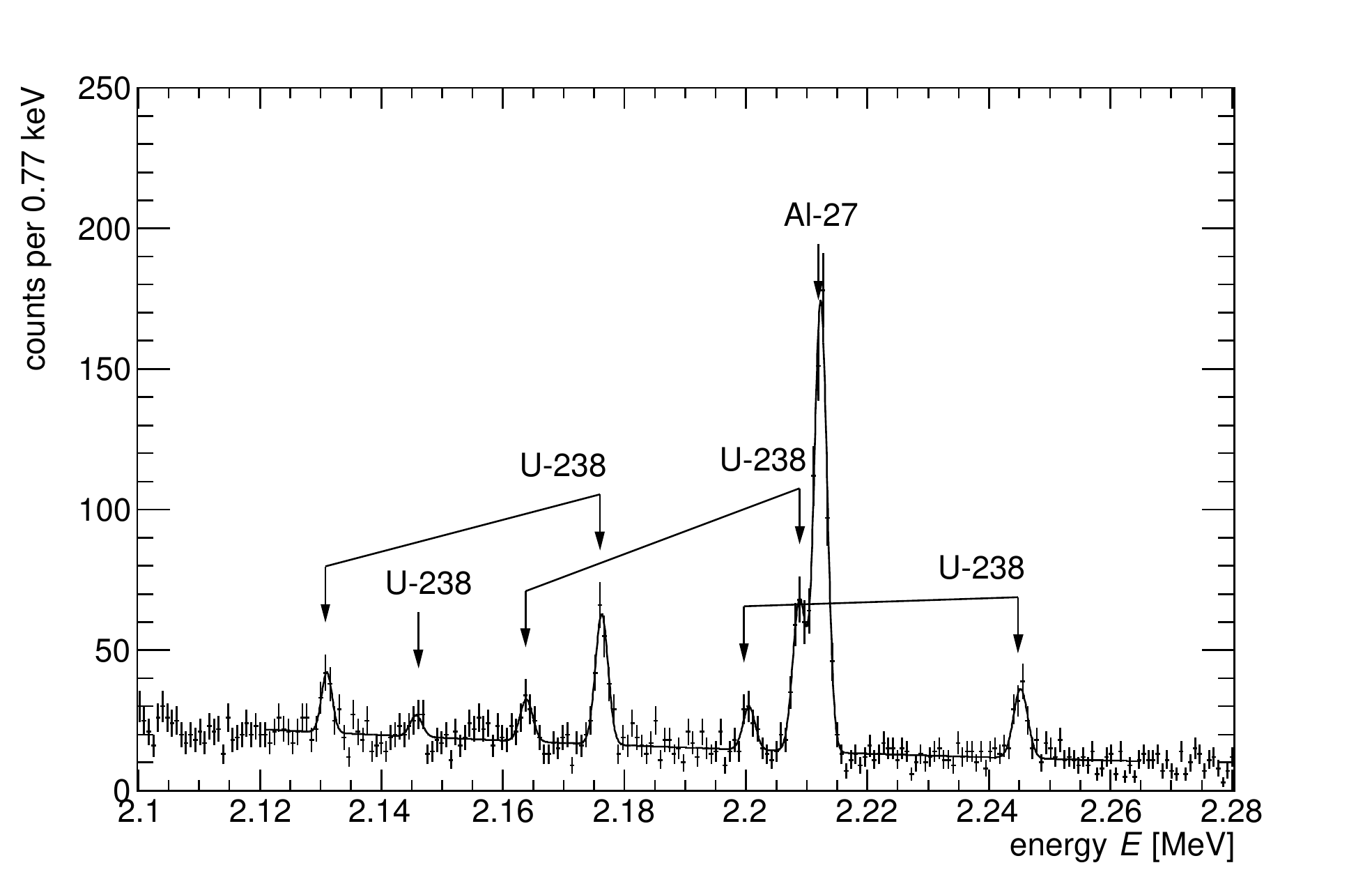}
\caption{A sample observed NRF spectrum in Detector 3 from Object 1, with corresponding fit (Eq.~\ref{eq:spectral_fit}, $\chi^2/\text{dof}=0.96$) overlain.}
\label{fig:det3_template1}
\end{figure}

In the signal region $2.12$--$2.26$~MeV, the observed spectra contain eight prominent peaks arising from $^{238}$U and $^{27}$Al NRF transitions. The $J^\pi = 1^+$ 2.176, 2.209, and 2.245~MeV levels in $^{238}$U exhibit transitions to the~$0^+$ ground state, producing peaks at these energies, and to the first $2^+$ state at 45~keV, producing the branched decay peaks 45~keV lower (see Fig.~\ref{fig:det3_template1}). $^{238}$U also contributes a small additional peak at 2.146~MeV, which is a non-branching ground state transition from the corresponding $1^-$ level. The large peak at 2.212~MeV arises from the 2.212~MeV $7/2^+$ state in  $^{27}$Al de-exciting to the $5/2^+$ ground state. Due to the non-zero energy resolution of HPGe detectors, these NRF peaks are observed as Gaussians with standard deviation $\mathcal{O}(\text{1 keV})$. These Gaussians sit atop an active background (well-described by a decaying exponential above $E \simeq 1$~MeV~\cite{hagmann2007photon}) resulting from a variety of non-resonant processes. 

Due to limited statistics, the observed spectra are first re-binned from the 32768 ADC channels to 4096 bins, giving bin widths of ${\lesssim}1$~keV. The spectra are next linearly calibrated based on the positions of the 2.212~MeV NRF peak, the 1.001~MeV passive DU line, and the 511~keV peak generated by pair production.
To compute the rate of NRF detections in each peak, the re-binned NRF spectra---see for example Fig.~\ref{fig:det3_template1}---are fit with Gaussian peaks atop a decaying-exponential continuum~\cite[Eq.~2]{vavrek2018experimental}:
\begin{align}\label{eq:spectral_fit}
f(E) = \exp\left( c_1 + c_2 E \right) + \sum_{k=1}^{8} \frac{a_k}{\sqrt{2\pi}\sigma_k} \exp\left[ - \frac{(E-E_k)^2}{2\sigma_k^2} \right],
\end{align}
where $a_k$ is the area of the $k^\text{th}$ peak, $\sigma_k$ is the standard deviation, and $E_k$ is the centroid energy.  The net count rate (above the continuum) under each peak is extracted as the peak area parameter $a_k$ divided by the spectrum bin width $\Delta E$~\cite[p.~171]{bevington2003error} and live charge $Q_\ell$, and the uncertainty is similarly $\delta n_\text{obs} = \delta a_k /\Delta E Q_\ell$, where $\delta a_k$ is the fit uncertainty in the area parameter $a_k$.

\section{Absolute NRF rate prediction}\label{sec:predictions}
It is possible to make absolute NRF rate predictions provided that quantities in the experimental design (such as the initial flux $\phi_0$ and various efficiency factors) and the underlying cross section parameters (for concreteness, we use the summary in Table~1 of Ref.~\cite{vavrek2018accuracy}) are known to good accuracy. Given the complexity of the experimental geometry in Section~\ref{sec:experiments}, the most accurate model for absolute NRF rates would be a full Geant4+G4NRF simulation starting from the electron beam striking the radiator and producing bremsstrahlung, continuing with photon transport in the measurement object and foil, and ending with photon energy deposition in the HPGe detectors. However, such an end-to-end model would be prohibitively computationally expensive (especially when running it repeatedly for different measurement objects), making it difficult to achieve low statistical uncertainty in the computed NRF rates. Instead, the full model can be split into various sub-models to pre-compute the effects of constant sections of the geometry (e.g., the radiator and foil) apart from any components that may change run-to-run (i.e., the measurement objects).

The first calculation split off from the hypothetical end-to-end model is the Geant4 simulation of the initial bremsstrahlung flux $\phi_0$ from electron interactions in the radiator. Combining the sampling of this distribution with a full photon transmission simulation from the exit of the radiator to energy deposition in the HPGe detectors will be referred to as \textbf{Model~A}.

Photon transmission from the foil to the detectors can also be split off from Model~A. This amounts to another separate Geant4 simulation that calculates the total detection probability for NRF photons emitted from the foil. Combining the bremsstrahlung simulation, photon transmission to the foil, and this pre-computed emission from the foil to the detectors results in a three-part \textbf{Model~B}. The following sections provide further detail on the model calculations.

In addition, the photon transmission and/or detection probability simulations can be replaced with semi-analytical models~\cite{vavrek2018accuracy} to greatly reduce computational time at a small expense to accuracy. Although we use these semi-analytical models in some of the later analysis, we are primarily concerned with validating the Geant4+G4NRF simulation against data, and thus we do not include details of the semi-analytical models in this work but instead refer the reader to Chapter~5 of Ref.~\cite{vavrek2019development}.

\subsection{Initial bremsstrahlung flux $\phi_0$}\label{sec:brems}
Common to both Model~A and B is a standalone Geant4 simulation of the $2.52$~MeV electron beam interaction with a high-fidelity model of the bremsstrahlung radiator to produce the initial flux $\phi_0$ as a function of energy $E$ and emission angle $\theta_b$ with respect to the electron beam axis. For computational efficiency, $\phi_0(E, \theta_b)$ is determined to high accuracy only in the regions of $E$-$\theta_b$ phase space that are important to NRF signal production, i.e., the photons with $E > 2$~MeV and angles $\theta_b$ that exit the collimator and reach the foil. This is achieved by tallying the $E$ and $\theta_b$ of photons that reach the plane of the foil in a simulation with no measurement object, and using this information to project back to a two-dimensional histogram $\phi_0(E, \theta_b)$ at the radiator. In subsequent simulations, the resulting $\phi_0(E, \theta_b)$ distribution is then sampled directly before the collimator entry. For further computational efficiency, the $\phi_0(E, \theta_b)$ distribution is not sampled directly, but through an importance sampling scheme. Photon energies are uniformly randomly sampled within $\pm 50$~eV of the resonances of interest (and angles sampled slightly wider than the collimator geometry), with the photons given a weight according to $\phi_0(E,\theta_b)$ to correct for the oversampling.

In all models, the raw absolute bremsstrahlung flux per incident electron (approximately $300$~photons/(eV$\cdot${\textmu}C) at 2.2~MeV) is further adjusted for the electron backscatter coefficient, which takes into account the fraction of electrons that scatter off the radiator into a component of the beamline that is electrically isolated from the charge collection area, and thus are not available for bremsstrahlung production. For the 126~{\textmu}m Au radiator and the 2.521~MeV beam, a separate Geant4 backscatter simulation of charge deposition in the radiator indicates that $0.932$~{\textmu}C are measured for every 1~{\textmu}C of incident electrons. Since the experimentally-obtained NRF rates are given in units of counts/{\textmu}C (measured), the input fluxes for the semi-analytical model or simulated with Geant4+G4NRF are divided by $0.932$ for consistency.

\begin{figure}[!h]
	\centering
	\includegraphics[width=0.8\columnwidth]{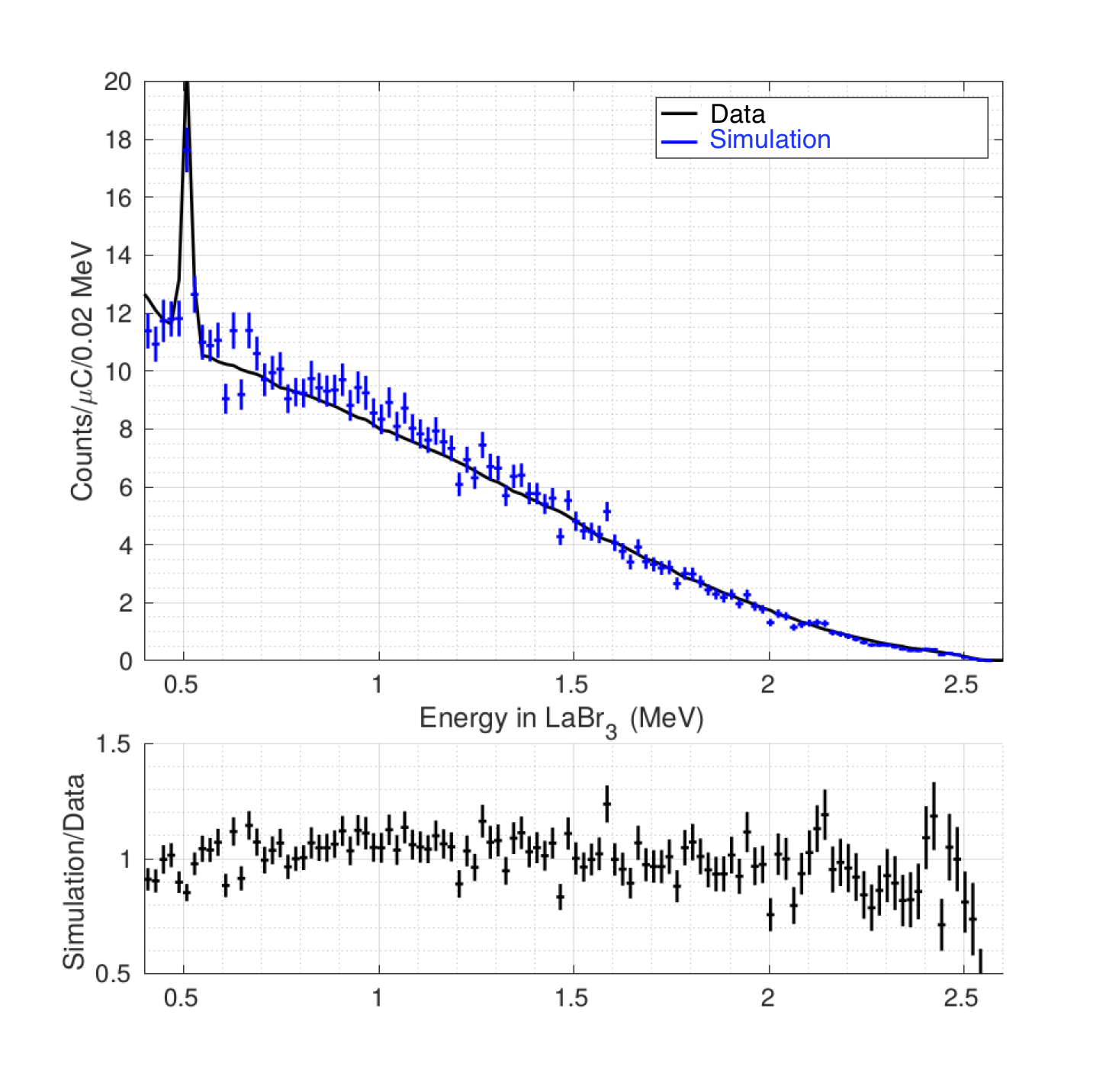}
	\caption{Observed (black) and simulated (blue) bremsstrahlung photon energy spectra in the LaBr$_3$ detector.
	The simulation includes scattering due to NRF and smearing due to detector resolution.  The comparison shows a good agreement, indicating that the bremsstrahlung flux is modeled correctly.}
	\label{fig:labr3_ThinGenu}
\end{figure}

Geant4 simulations of the expected counts in the downstream LaBr$_3$ detector provide evidence that the upstream components---i.e., the bremsstrahlung output and the object and foil geometries---are modeled correctly. The ratio of counts observed by the LaBr$_3$ scintillator to the simulated counts is overall consistent with unity.  Fig.~\ref{fig:labr3_ThinGenu} shows plots of experimental data with the results of the Monte Carlo simulation overlain, and the ratio of observed and simulated counts bin-by-bin.  The comparison shows a strong agreement, in particular in the most relevant 1.9-2 MeV energy domain. Furthermore, the modeled LaBr$_3$ efficiency was found to be in good agreement with data taken using calibrated $^{60}$Co and $^{137}$Cs sources. 

\subsection{Photon transport to the foil and detectors}
The photons generated in Section~\ref{sec:brems} are injected into the simulation geometry directly between the radiator exit and collimator entry. Photons transmitted through the collimator then impinge on the simulated measurement object, where they can interact non-resonantly through standard Geant4 electromagnetic processes (G4EmStandardPhysics, with default options) or resonantly through G4NRF. The transmitted flux $\phi_t(E)$ then impinges on the foil. In Model~A, this flux is allowed to undergo NRF interactions in the foil and travel back towards the HPGe detectors where the number of full-energy depositions is tallied. These simulations required roughly $10^{11}$ importance-sampled primary photons (${\sim}15\,000$~cpu-hours) to obtain statistical uncertainty of ${<}5\%$ in the thickest geometry (Object~3), though runtimes here could potentially be further improved by adjusting the importance sampling scheme.

In Model~B, conversely, the flux striking the foil is allowed to produce NRF photons, but information about these photons is recorded and the event is ended early instead of letting the photons exit the foil towards the HPGe detectors. This simulation thus computes the NRF production rate in the foil, which can then be multiplied by the total detection probability to compute the rate of NRF detections in each detector. The total detection probability $P_d(E')$ is defined as the probability for a photon of energy $E'$ generated in the foil to escape the foil, be transmitted towards detector $d$ and through the Pb filter, and deposit its full energy in the active volume of the HPGe crystal. As such, it accounts for both geometric and intrinsic detector efficiencies, as well as losses due to any filtering material between the foil and detector. $P_d(E')$ decreases exponentially with depth $x$ into the foil, due to attenuation from the foil itself:
\begin{align}\label{eq:detection_prob}
P_d(E', x) = \beta_0 \exp\left(-\beta_1 x\right),
\end{align}
where $\beta_0$ and $\beta_1$ are constants for a given detector $d$ and energy $E'$.

To compute the detection probabilities, photons of various energies $E'$ are generated in the beam-illuminated portions of the DU and Al foil layers and allowed to propagate towards the HPGe detectors. The number of full-energy depositions is recorded, giving (for each detector) a probability of full-energy detection $P_d(E')$ for an NRF photon with energy $E'$ created at a depth $x$ in the foil. For a given detector and NRF line of interest, the histogram of detection probability vs initial $x$ in the foil is fit with Eq.~\ref{eq:detection_prob}, giving $\beta_0$ and $\beta_1$. These detection probability simulations required a single run of $8\times 10^9$ events (${\sim}800$~cpu-hours) to achieve a statistical uncertainty of ${\sim}0.01\%$ (for $^{238}$U) or ${\sim}1\%$ (for $^{27}$Al) in the fit parameters $\beta_0$ and $\beta_1$, after which only a further $1\times 10^9$ events (${\sim}100$~cpu-hours) in each of the photon propagation simulations was required for an NRF rate uncertainty of ${\sim}0.3\%$. These statistical uncertainties are much smaller than those from the observed data, and are excluded from error propagation calculations.

Both Model~A and B require high-fidelity Geant4 models of the HPGe detector geometries. In particular, the dead layer of germanium surrounding the active volume can substantially reduce detector efficiency~\cite{rodenas2003deadlayer} through two compounding effects: not only is the active volume of the detector reduced, but a layer of shielding is effectively added around the detector as well. Comparisons of separate Geant4 efficiency simulations against data taken with a $^{137}$Cs calibration source indicate that the manufacturer-provided dead layer specifications (in this case, $0.7$~mm) of our HPGe detectors are inaccurate. Dead layer thicknesses for the detectors used in this work are estimated by comparing observed vs.~simulated counts using a $^{137}$Cs calibration source (see, e.g., Ref.~\cite{karamanis2002efficiency}), and varying the simulated dead layer thickness until the observed $662$~keV counts are best replicated.

\section{Results}\label{sec:results}
\subsection{Model-to-data comparisons}
Table~\ref{tab:avg_ratios_Er} shows the ratios of NRF count rates predicted by the two models over observed rates. The $^{238}$U and $^{27}$Al cross section parameters assumed in the models are summarized in Table~1 of Ref.~\cite{vavrek2018accuracy}, and are derived or taken from Refs.~\cite{quiter2011transmission, endt1990levels}---see Section~\ref{sec:discussion}. Ratios are grouped through a weighted average for each major ground-state transition ($E_r = 2.176$, $2.245$, and $2.212$~MeV). The $E_r = 2.209$~MeV line is excluded from the analysis due to its large uncertainties induced by the overlap with the $2.212$~MeV peak. 
Row averages over the two models for each grouping are computed in the final column of the table, using fully-correlated uncertainty propagation because the uncertainty in each column arises from fit uncertainty in the same dataset. 
Possible reasons for these discrepancies from unity are discussed in Section~\ref{sec:discussion}, and additional systematic uncertainties are discussed in Section~\ref{sec:systematics}.

\begin{table}[!h]
	\centering
	\begin{tabular}{c|c|c||c}
	$E_r$ [MeV] & Model A & Model B & avg \\\hline
	2.176 & $1.207 \pm 0.025$ & $1.203 \pm 0.024$ & $1.205 \pm 0.025$\\
2.245 & $1.158 \pm 0.038$ & $1.147 \pm 0.036$ & $1.152 \pm 0.038$\\
2.212 & $1.111 \pm 0.017$ & $1.068 \pm 0.016$ & $1.088 \pm 0.017$\\

	\end{tabular}
	\caption{Ratio of predicted NRF rates to observed NRF rates for the two models, averaged over the three detectors and all runs. 
	}
	\label{tab:avg_ratios_Er}
\end{table}

Equipped with both the absolute NRF count rate predictions and data, we can also perform the more typical relative analysis in which the count rate in an NRF line of interest is normalized against another (often better-understood) NRF line. This normalization has the benefit of eliminating (or at least substantially reducing) systematic factors (and their uncertainties) such as the absolute bremsstrahlung flux $\phi_0(E)$ and HPGe detection probabilities $P_d$. In Table~\ref{tab:avg_Rratios_det}, the quantity $R$ is formed by normalizing both the observed and predicted rates in each NRF line by another NRF line and dividing the two resulting ratios, e.g.:
\begin{align}\label{eq:R}
R_{2176/2212} \equiv \left(\frac{n_\text{2176~keV}}{n_\text{2212~keV}}\right)_\text{predicted} \bigg/ \left(\frac{n_\text{2176~keV}}{n_\text{2212~keV}}\right)_\text{observed}.
\end{align}
As above, if the underlying physics models in G4NRF are correct, the $R$ ratios should be consistent with unity. Table~\ref{tab:avg_Rratios_det} gives a list of $R$ values separated by detector and model, though separations over run date, e.g., are also possible. Averaged over the two models and all three detectors, the $R$ ratios for the three possible line pairs range from $0.96$--$1.03$, indicating only ${\lesssim}4\%$ deviations from unity (which are statistically indistinguishable from zero) in the relative NRF rates.

\begin{table}[!h]
	\centering
	\begin{tabular}{c|c|c|c||c}
	Detector & $R$ & Model A & Model B & avg \\\hline\hline
	0   & 2176/2245 & $0.905 \pm 0.063$ & $0.939 \pm 0.064$ & $0.922 \pm 0.063$\\
    & 2176/2212 & $0.995 \pm 0.046$ & $1.054 \pm 0.047$ & $1.024 \pm 0.047$\\
    & 2245/2212 & $0.998 \pm 0.065$ & $1.031 \pm 0.065$ & $1.015 \pm 0.066$\\\hline
2   & 2176/2245 & $1.018 \pm 0.074$ & $1.007 \pm 0.070$ & $1.012 \pm 0.070$\\
    & 2176/2212 & $1.070 \pm 0.054$ & $1.098 \pm 0.053$ & $1.084 \pm 0.055$\\
    & 2245/2212 & $0.967 \pm 0.066$ & $0.980 \pm 0.065$ & $0.974 \pm 0.067$\\\hline
3   & 2176/2245 & $0.972 \pm 0.063$ & $0.951 \pm 0.059$ & $0.961 \pm 0.059$\\
    & 2176/2212 & $0.988 \pm 0.049$ & $1.008 \pm 0.048$ & $0.998 \pm 0.050$\\
    & 2245/2212 & $0.997 \pm 0.063$ & $1.034 \pm 0.063$ & $1.015 \pm 0.065$\\\hline\hline
avg & 2176/2245 & $0.960 \pm 0.038$ & $0.963 \pm 0.037$ & $0.962 \pm 0.037$\\
    & 2176/2212 & $1.013 \pm 0.029$ & $1.051 \pm 0.028$ & $1.032 \pm 0.029$\\
    & 2245/2212 & $0.988 \pm 0.038$ & $1.015 \pm 0.037$ & $1.002 \pm 0.038$\\

	\end{tabular}
	\caption{Predicted-to-observed ratios $R$ of ratios (i.e., relative validation results) of NRF lines (Eq.~\ref{eq:R}) within the same spectrum.}
	\label{tab:avg_Rratios_det}
\end{table}

\subsection{Systematic uncertainties}\label{sec:systematics}
The uncertainties discussed so far have primarily been uncertainties of an ultimately statistical origin, either in the brute-force Model~A simulation results or the uncertainty of the fit to Eq.~\ref{eq:spectral_fit}. Additional systematic uncertainties in the absolute NRF rates can be broken into two primary categories: quasi-random (but not statistical) variations that are uncorrelated across the observed rates, and constant biases that affect some or all of the rates in the same way. In the absence of detailed probability distributions, uncertainties of the former category are modeled with normal distributions $p_X \sim \mathcal{N}(\mu,\sigma)$, while the latter are modeled with uniform distributions $p_X \sim \text{U}(x_1, x_2)$. A list of major contributions to the systematic uncertainty estimation is as follows:

\begin{enumerate}
\item \textbf{data-wide systematic variations}: we first test whether the statistical uncertainties of the 21 statistically-independent model-to-data ratios for each of the three NRF lines are sufficiently large to account for the observed variation in each line's dataset. Using the set of ratios predicted through Model~B, the reduced $\chi^2$ statistics for the $\{2.176, 2.245, 2.212\}$~MeV datasets are approximately $\{2.4,1.9,2.7\}$ and have $p$-values on the order of $\{10^{-3}, 10^{-2}, 10^{-5}\}$, indicating that the statistical uncertainties are not large enough to explain the weighted sample standard deviations of $\{0.17, 0.23, 0.12\}$. We therefore conclude that there are additional random but systematic uncertainties in the means of each line's dataset---on top of statistical fluctuations---due to possible effects such as beam flux variations not captured by the current readout
. Assuming these fluctuations are roughly normally-distributed, the total uncertainties on the means are estimated as $\{0.17, 0.23, 0.12\}/\sqrt{21} = \{ 0.036, 0.050, 0.026\}$. Subtracting the Model~B statistical uncertainties in quadrature (cf.~Table~\ref{tab:avg_ratios_Er}), the random systematic components are $\{0.027, 0.034, 0.020 \}$.

\item \textbf{intrinsic detector efficiencies}: the activity of the $^{137}$Cs source used to conduct the efficiency calibration has two uncertainties distributed as $\mathcal{N}(0, 3\%)$ and $\text{U}(-5\%,5\%)$, arising from a comparison against another $^{137}$Cs source that was guaranteed to a tolerance of $\pm 5\%$ by the manufacturer (Spectrum Techniques). Furthermore, the efficiency correction for the nominal dead layer thickness is uncertain to about $\pm 1.5\%$. Although this latter correction varies somewhat with detector number, for simplicity we assign the most conservative uncertainty and include a distribution of $\mathcal{N}(0, 1.5\%)$ for all three detectors.

\item \textbf{pileup}: based on an analysis of unrejected pileup in the NRF spectral data, we estimate that a maximum of $2.5\%$ of true NRF peak events are lost due to pileup with lower-energy photons, and therefore include a systematic uncertainty distributed as $\text{U}(0,2.5\%)$ in the denominators of the NRF model-to-data ratios in Section~\ref{sec:results}.

\item \textbf{notch refill}: calculations of the transmitted flux $\phi_t(E)$ do not account for the \textit{notch refill} effect: the downscattering of higher energy photons to the resonance energy $E_r$ via small-angle Compton scattering in the measurement object. Model~A and B both neglect notch refill due to the resonance importance sampling scheme. Simple analytical estimates~\cite{pruet2006detecting} suggest that notch refill would increase the modeled on-resonance flux by an estimated ${\sim}0.7\%$ in the thinner Objects 1--2 and ${\sim}1.5\%$ in the thicker Objects 3--5. Unlike the pileup uncertainty, this is not a strict bound, so we include a single uniformly-distributed uncertainty with $1.5\%$ as its midpoint, i.e., $\text{U}(0,3\%)$, in the numerators of the model-to-data ratios.

\end{enumerate}
The net effect of these uncertainties is computed through a Monte Carlo calculation (assuming all uncertainties are uncorrelated) for each of the three NRF lines studied. The final uncertainty distributions are still approximately Gaussian, but the means of the distributions are slightly shifted from unity to $1.002$--$1.003$ due to the asymmetric contributions of the uniform distributions. Neglecting this small shift, the $\pm 34\%$ confidence intervals around unity correspond to approximately $\pm 5\%$ relative fluctuations around the mean, for final model-averaged predicted over observed count rate ratios in each NRF line (cf.~Table~\ref{tab:avg_ratios_Er}) of

\begin{align*}
\text{$^{238}$U, 2.176 MeV: }& 1.208 \pm 0.025 \text{ (stat.) } \pm 0.061 \text{ (sys.)}\\
\text{$^{238}$U, 2.245 MeV: }& 1.154 \pm 0.038 \text{ (stat.) } \pm 0.062 \text{ (sys.)}\\
\text{$^{27}$Al, 2.212 MeV: }& 1.090 \pm 0.017 \text{ (stat.) } \pm 0.054 \text{ (sys.)}
\end{align*}

Combining the statistical errors 
linearly with the systematic errors, deviations from unity in these ratios are found at levels of $+2.4$, $+1.5$, and $+1.3$ standard deviations, respectively, indicating that the assumed NRF cross section parameters (Table~1 of Ref.~\cite{vavrek2018accuracy}) are approximately consistent with the observed rates.

\subsection{NRF cross section parameters}
Rather than compare observed and predicted count rates using a set of assumed NRF cross section parameters, we can instead invert the analysis and use the measured NRF rates to infer the NRF cross section parameters. As mentioned, the analysis (especially of uncertainties) is complicated by the use of multiple different run configurations, but we include the results here for completeness.

For each observed spectrum (see, e.g., Fig.~\ref{fig:det3_template1}), we first extract each pair of $^{238}$U branching ratios for both the 2.176 and 2.245~MeV levels. 
Since the $^{238}$U branched decays are only $45$~keV lower in energy than their corresponding ground state decays, the difference in non-resonant attenuations, intrinsic efficiencies, and filter transmission probabilities is only ${\sim} 1\%$ and can be neglected to good approximation. The $W(\theta,\phi)$ for the branched decays is determined by the $0\to 1\to 2$ spin sequence rather than the $0\to 1\to 0$ sequence of the ground state decay, but the ratio of the two $W(\theta,\phi)$ at $\theta_d \simeq 125^\circ$ is very close to unity. The ratio of observed counts in the ground state vs branched decay peaks is therefore very nearly equal to the ratio of branching ratios for each level~$r$:
\begin{align}\label{eq:br_extraction}
\frac{n_\text{g.s.}}{n_\text{br}} = \frac{b_{r,0}}{b_{r,1}},\, \text{ with } b_{r,0} + b_{r,1} = 1.
\end{align}
Eq.~\ref{eq:br_extraction} therefore contains two equations and two unknowns, and thus can be used to provide an estimate of $b_{r,0}$ and $b_{r,1}$ for each $^{238}$U level $r$ in each spectrum, independent of the value of $\Gamma_r$. The resulting weighted averages are:
\begin{align}
\label{eq:br0_inferred_first}
E_r = 2.176\text{ MeV: } b_{r,0} = 0.676 \pm 0.010\text{ (stat.)}\\
E_r = 2.245\text{ MeV: } b_{r,0} = 0.649 \pm 0.013\text{ (stat.)}
\label{eq:br0_inferred_last}
\end{align}
where the systematic uncertainties have canceled by forming the ratio in Eq.~\ref{eq:br_extraction}.

Given these experimentally-determined branching ratios, we use the semi-analytical models described in Section~\ref{sec:predictions} to tabulate predicted absolute NRF rates as a function of level width $\Gamma_r$, since it is much faster to re-evaluate the semi-analytical models for arbitrary $\Gamma_r$ than are the simulations of Models~A or B. The tables of rates vs widths---one table for each detector, measurement object, and NRF line---are then interpolated using observed NRF rates to obtain experimentally-determined values of $\Gamma_r$ for the three NRF lines of interest (a similar technique was used in, e.g., Fig.~3 of Ref.~\cite{pietralla1995absolute}). Once again performing a weighted average over all detectors and run dates, the inferred NRF level widths are:
\begin{align}
\label{eq:Gammar_inferred_first}
E_r = 2.176\text{ MeV: } \Gamma_{r} = 37.3~\pm 1.0 \text{ (stat.) } \pm 3.3 \text{ (syst.)~meV}\\
\label{eq:Gammar_inferred_second}
E_r = 2.245\text{ MeV: } \Gamma_{r} = 22.7~\pm 0.9 \text{ (stat.) } \pm 1.6 \text{ (syst.)~meV}\\
E_r = 2.212\text{ MeV: } \Gamma_{r} = 14.8~\pm 0.3 \text{ (stat.) } \pm 1.0 \text{ (syst.)~meV}
\label{eq:Gammar_inferred_last}
\end{align}

The most probable values of $\Gamma_r$ above are calculated through a weighted average of the inferred $\Gamma_r$ from each experimental run. The weights for each $\Gamma_r$ are determined by the (independent) statistical uncertainties in each run's inferred $\Gamma_r$ in order to estimate both a final, average value of $\Gamma_r$ and its statistical uncertainty separate from correlated systematic uncertainties. 


\section{Discussion and Conclusions}\label{sec:discussion}
The experiments of Ref.~\cite{vavrek2018experimental}  provide data for an improved G4NRF validation study---in terms of both absolute and relative measurements---over previous results. Ref.~\cite{jordan2007simulation}, which introduced the initial G4NRF code on which our updated version~\cite{vavrek2018accuracy} is based, saw a factor of ${\sim}3$ discrepancy in absolute NRF rates~\cite{warren_pers_comm} and a factor of ${\sim}30\%$ in relative rates. The independent Geant4 NRF code of Ref.~\cite{lakshmanan2014simulations} could not be validated against absolute measurements, but was validated to within $20\%$ in relative NRF rates. In addition, our results allow for the measurement of absolute NRF cross sections, rather than relying on normalization via a separate $^{27}$Al target or the $511$~keV peak.  To ascertain that the incident bremsstrahlung flux is modeled correctly, data from a downstream LaBr$_3$ detector is compared to separate beam transport simulations, showing a strong agreement.

As shown in Section~\ref{sec:systematics} above, this study observes an agreement to within 1.3-2.4 standard deviations between observed experimental data and modeling based on G4NRF which uses existing NRF cross sections. 
Grouping by detector
, a moderately smaller discrepancy factor of $1.04$ is observed in Detector 2 as compared to factors of $1.17$ and $1.19$ in Detectors~0 and 3 -- this variation of results is part of the systematic uncertainties, as reported in Equations \ref{eq:Gammar_inferred_first}-\ref{eq:Gammar_inferred_last}. 
It should also be pointed out that most NRF level widths in past literature were extracted in experiments which normalized to the $^{27}$Al line.  If we normalize the results in Section~\ref{sec:systematics} to the $^{27}$Al result, then the discrepancy for the $^{238}$U lines become only 1.4 and 0.6 standard deviations.
Table~\ref{tab:gamma_br} lists the extracted level widths and branching ratios along with results previously reported in the literature.  It can be observed that any of the discrepancies from past results are often on the order of ---and in some cases within---the variations in the literature itself.  The  $^{27}$Al inferred level width of $\Gamma_r = 14.7$~meV 
agrees with literature values of the level width of $\Gamma_r = 17.1$~meV in Table~27.4 of Endt~\cite{endt1990levels} or $\Gamma_r = 17.29$~meV in Pietralla et al.~\cite{pietralla1995absolute} at the level of ${\sim}15\%$ (see Table~\ref{tab:gamma_br}).  The branching ratios extracted by this study agree to withing statistical uncertainties and the variability in the past reported values.



\begin{table}[!htb]
\centering
\begin{tabular}{r|cccc|c}
Reference & $\Gamma_{2176}$~[meV] & $b_{2176,0}$ & $\Gamma_{2245}$~[meV] & $b_{2245,0}$ & $\Gamma_{2212}$~[meV] \\\hline\hline
this work                                         & 37.3 $\pm$ 4.3 & 0.676 $\pm$ 0.010 & 22.7 $\pm$ 2.5 & 0.649 $\pm$ 0.013 & 14.7 $\pm$ 1.3 \\\hline
ENSDF~\cite{nndc2015u238,nndc2011al27}            & 58.0 & 0.649 & 1.42 & 0.877 & 17.2 \\\hline
Hammond et al.~(I)~\cite{hammond2012dipole}       & 56.6 & 0.651 & 18.5 & 0.876 & --\\
Quiter et al.~(1)~\cite{quiter2011transmission}   & 54.8 & 0.658 & 29.0 & 0.680 & --\\
Quiter et al.~(6)~\cite{quiter2011transmission}   & 47.4 & 0.645 & 35.2 & 0.658 & --\\
Heil et al.~(3)~\cite{heil1988observation}        & 58.1 & 0.658 & 33.5 & 0.680 & --\\\hline
Pietralla et al.~(I)~\cite{pietralla1995absolute} & -- & -- & -- & -- & 17.29 \\
Endt et al.~(27.4)~\cite{endt1990levels}          & -- & -- & -- & -- & 17.1 \\

\end{tabular}
\caption{Summary of level widths $\Gamma_r$ and ground-state branching ratios $b_{r,0}$ from this work, previous experiments, and the ENSDF evaluation.  The reported errors on the level width are the sums of systematic and statistical uncertainties. For experimental papers, the relevant table number is given in parentheses in the first column.}
\label{tab:gamma_br}
\end{table}



The relative rate analysis---see the $R$ ratios given in Eq.~\ref{eq:R} and  Table~\ref{tab:avg_Rratios_det}---further indicates that the NRF physics is modeled correctly to within ${\lesssim}4\%$, as experimental ratios of two NRF lines overall agree with the G4NRF model predictions to within experimental uncertainties. In fact, the near-unity values of $R$ in Table~\ref{tab:avg_Rratios_det} suggest that a relative cross section measurement in which the $^{238}$U NRF rates were normalized to those of the $^{27}$Al 2.212~MeV line would return $^{238}$U cross section parameters in greater agreement with their assumed values. Moreover, the near-unity values of the $^{238}$U to $^{27}$Al ratios suggests that discrepancies in the absolute results are not due to any residual polarization of the bremsstrahlung beam, which would affect the $^{238}$U and $^{27}$Al rates very differently due to their different $W(\theta,\phi)$.


We have thus validated the G4NRF code for Geant4 against both absolute and relative measurements of $^{238}$U and $^{27}$Al NRF count rates from data obtained in a previous experiment~\cite{vavrek2018experimental}. These results provide, based on absolute measurements, experimental validation of G4NRF at a level of $8\%$ in $^{27}$Al and $15$--$20\%$ in $^{238}$U, compared to statistical errors of $2$--$3\%$ and systematic errors of approximately $5\%$. The level of validation in the relative rate analysis, by contrast, is ${\lesssim}4\%$, and is statistically indistinguishable from zero. Such agreement between the two Geant4+G4NRF count rate models and the observed NRF rates provides good predictive capability for absolute NRF count rates in real measurements.  Inverting the validation analysis, we also determined the transition widths of the various NRF transitions, with results that are within 15-20 \% of previously reported values.

\section*{Acknowledgements}
This work was funded in-part by the Consortium for Verification Technology under Department of Energy National Nuclear Security Administration award number DE-NA0002534. JRV gratefully acknowledges the support of the Nuclear Regulatory Commission grant number NRC-HQ-84-15-G-0045, and BSH gratefully acknowledges the support of the Stanton Foundation's Nuclear Security Fellowship program. The authors wish to thank the Massachusetts Green High Performance Computing Center (MGHPCC) for computing resources, as well as Richard Milner for comments on the manuscript.

\bibliographystyle{unsrt}
\bibliography{validationpaper.bib}

\end{document}